\documentclass[prl,twocolumn,showpacs,floatfix,amsfonts]{revtex4}
\usepackage{graphicx,graphics,color,epsfig}
\usepackage{bm}
\usepackage{amsmath}
\usepackage{amssymb}
\usepackage{epstopdf}
\begin{document}
\preprint{}
\title{Fermi surface and antiferromagnetism in the Kondo lattice:\\ 
an asymptotically exact solution in $d>1$ dimensions}
\author{Seiji J. Yamamoto and Qimiao Si}
\affiliation{Department of Physics \& Astronomy, Rice University,
Houston, TX 77005, USA }
\begin{abstract}

Interest in the heavy fermion metals has motivated us to examine 
the quantum phases and their Fermi surfaces
within the Kondo lattice model. We demonstrate that
the model is soluble
asymptotically exactly in any dimension $d>1$, 
when the 
Kondo coupling 
is small compared with the RKKY interaction
and in the presence of antiferromagnetic ordering.
We show that 
the Kondo coupling is exactly marginal in
the renormalization group sense,
establishing 
the stability 
of an ordered phase with a small Fermi surface, ${\rm AF_S}$.
Our results have implications for the global phase diagram of the
heavy fermion metals, suggesting 
a Lifshitz transition inside the antiferromagnetic 
region and providing a new perspective for a Kondo-destroying 
antiferromagnetic quantum critical point.

\end{abstract}
\maketitle

\narrowtext

There is a growing list of materials in which quantum criticality may have 
a strong influence on their electronic and magnetic properties.
A basic question is whether 
quantum criticality
can be adequately described
in terms of order-parameter fluctuations,
or if
inherently quantum effects must be incorporated.
It is fortunate that 
the issue 
can be 
systematically 
studied in heavy fermion metals~\cite{stewart2001, HvL2006}.
These systems involve 
the Kondo effect which, via the Kondo singlet formation,
is an inherently quantum property. Indeed, the discussion 
here 
has centered on whether or not the Kondo entanglement
energy scale collapses at the antiferromagnetic quantum critical
point (QCP). If the Kondo scale remains finite, the QCP is of the 
spin-density-wave (SDW) type~\cite{hertz1976, millis1993, moriya1985}.
If it collapses, new types of theory~\cite{si2001, si2003, coleman2001, senthil2004a}
-- such as the local quantum criticality  -- are needed.

To understand the QCPs, it is instructive
to elucidate the proximate quantum phases.
The ordered phase involved in
the SDW QCP is expected to be an antiferromagnet whose Fermi surface 
incorporates both the conduction electrons and local moments;
such a Fermi surface is called ``large'' and the corresponding phase
is named ${\rm AF_L}$.
For a Kondo-destroying QCP, on the other hand, it must be
an ${\rm AF_S}$ phase, whose Fermi surface is ``small''
in the sense that it encloses only the conduction electrons. 
Experimentally, evidence exists from dHvA measurements for 
the ${\rm AF_S}$ phase~\cite{onuki2003, mccollam2005}
and, moreover,
there are also indications~\cite{paschen2004, shishido2005} from 
Hall-effect and dHvA for a direct transition from
the ${\rm AF_S}$ phase to ${\rm PM_L}$,
a paramagnetic metal phase with a large Fermi surface.
Theoretically, however, whether the ${\rm AF_S}$ is 
a stable phase of the Kondo lattice with spin-rotational invariance
has not been previously established in dimensions higher than one.
In this letter,
we answer the question in the affirmative for the
model with SU(2) spin symmetry.
Our results
are asymptotically exact, something that is ordinarily difficult to
achieve for any correlated-electron model in more than one
dimension.

We consider the Kondo lattice model:
\begin{eqnarray}
	\mathcal{H} &=& \mathcal{H}_f +
\mathcal{H}_c +\mathcal{H}_K 
\end{eqnarray}
Here, $\mathcal{H}_c = \sum_{\vec{k}\sigma}\epsilon_{\vec{k}}
\psi^{\dagger}_{\vec{k}\sigma}
\psi_{\vec{k}\sigma}$ describes a band of free conduction $c-$electrons, 
with a bandwidth $W$.
For simplicity, we will consider the electron concentration, $x$ per site,
to be
such that the Fermi surface
of $\mathcal{H}_c$ alone
does not touch the antiferromagnetic zone boundary.
$\mathcal{H}_K = \sum_{i} J_K {\vec S}_i \cdot {\vec s}_{c,i}$ 
specifies the
Kondo interaction of strength $J_K$;
here the conduction electron spin $\vec{s}_{c,i} = \frac{1}{2}
\sum_{\sigma\sigma^{\prime}}\psi^{\dagger}_{\sigma,i} 
\vec{\tau}_{\sigma\sigma^{\prime}}\psi_{\sigma^{\prime},i}$, 
where $\vec{\tau}$ is the vector of Pauli matrices.
Finally, $\mathcal{H}_f = \frac{1}{2}
\sum_{ij} I_{ij} {\vec S}_i \cdot {\vec S}_j$
is the magnetic Hamiltonian for the spin-$\frac{1}{2}$
$f-$moments, ${\vec S}_i$,
for which there is $1$ per site. The strength of the exchange interactions,
$I_{ij}$, is characterized by, say, the nearest neighbor value, $I$. 

\begin{figure}[t]
   \centering
   \includegraphics[width=2.0in]{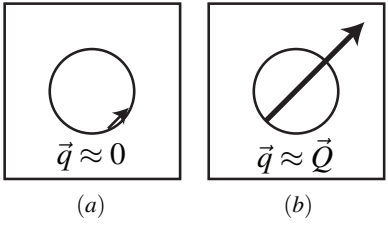}
   \caption{With the Fermi surface (FS) of the conduction-electron component not touching
the antiferromagnetic zone boundary, only the uniform component ($\vec{q} \approx 0$) of the local moments
can interact with two states near the FS, as shown in (a).  The linear
coupling involving the staggered component, $\vec{n}\cdot\vec{s}_c$, is not kinematically 
favorable, as shown in (b).}
   \label{fig:fs-match}
\end{figure}

{\it QNL$\sigma$M representation of the Kondo lattice:~}
We focus on the parameter region with $J_K \ll I \ll W$. Here, it is 
appropriate to expand around the limit $J_K=0$, where the local-moment and 
conduction-electron components are decoupled.
We will consider, for simplicity, square or cubic lattices, although
our results will be generally valid provided that the ground 
state
is a
collinear antiferromagnet.
$\mathcal{H}_f$ can be mapped to a quantum non-linear sigma model
(QNL$\sigma$M) by standard means \cite{haldane1983, chakravarty1989}.
The low-lying excitations are concentrated in the momentum space near
${\vec q}={\vec Q}$ (the staggered magnetization) and 
near ${\vec q}={\vec 0}$ 
(the total magnetization being conserved):
\begin{equation}
	2 \vec{S}_i 
\to \eta_{\vec{x}} \vec{n}(\vec{x},\tau)\sqrt{1-\left( 
2a^d \vec{L}(\vec{x},\tau) \right)^2} 
+ 
2a^d
\vec{L}(\vec{x},\tau) 
\end{equation}
where $\vec{x}$ labels the position,
$\eta_{\vec{x}} = \pm 1$ on even and odd sites,
and $a$ is the lattice constant.
The linear coupling $\vec{n}\cdot\vec{s}_c $ cannot connect two 
points on the Fermi surface and is hence unimportant for low energy
physics (such a kinematic constraint has appeared in other
contexts, {\it e.g.} Ref.~\cite{sachdev95});
see Fig. \ref{fig:fs-match}b. 
The Kondo coupling is then replaced by an effective one,
$\vec{S} \cdot\vec{s}_c \rightarrow 
a^d
\vec{L}\cdot\vec{s}_c$,
corresponding to forward scattering for the conduction electrons;
see Fig. \ref{fig:fs-match}a.

The mapping to the QNL$\sigma$M can now be implemented by
integrating out the $\vec{L}$
field. The effective action 
is
\begin{eqnarray}
	\mathcal{S} &=& \mathcal{S}_{\text{QNL}\sigma\text{M}}+\mathcal{S}_{\text{Berry}}+\mathcal{S}_K+\mathcal{S}_c\\
	\mathcal{S}_{\text{QNL}\sigma\text{M}} &\equiv& \frac{c}{2g}\int d^dxd\tau\left[ \left(\nabla \vec{n}(\vec{x},\tau)\right)^2 + 
\left(\frac{\partial\vec{n}(\vec{x},\tau)}{ c ~\partial\tau}\right)^2 \right] 
\nonumber \\
	\mathcal{S}_K &\equiv& 
\lambda\int d^dx d\tau\left[ \vec{s}_c(\vec{x},\tau)\cdot \vec{\varphi}(\vec{x},\tau) \right] \nonumber \\
	\mathcal{S}_c &\equiv& \int d^d K d\varepsilon 
\sum_{\sigma} \psi^{\dagger}_{\sigma}(\vec{K},i\varepsilon) 
\left(i\varepsilon - \xi_{K} \right)\psi_{\sigma}(\vec{K},i\varepsilon) \nonumber\\
	&& + \lambda^2\int \psi^4
\nonumber
\end{eqnarray}
where $\xi_K \equiv v_F(K-K_F)$.
The Berry phase term for the
$\vec{n}$ field,
$\mathcal{S}_{\text{Berry}}$, 
is not important
inside the N\'{e}el phase.
We have introduced 
a vector boson field $\vec{\varphi}$ which 
is shorthand 
for
$\vec{n}\times\frac{\partial\vec{n}}{\partial\tau}$.  
The $\vec{n}$ field satisfies the constraint, $\vec{n}^2=1$, which is solved by
$\vec{n} = (\vec{\pi}, \sigma)$, where 
$\vec{\pi}$ labels the Goldstone magnons
and $\sigma \equiv \sqrt{1-\vec{\pi}^2}$ is the massive field.
We will consider the case of a
spherical Fermi surface; since only forward scattering is important,
our results will apply 
for more complicated Fermi-surface geometries.
The parameters for the QNL$\sigma$M 
will be considered as phenomenological \cite{chakravarty1989},
though they can be explicitly written in terms of the 
microscopic parameters.
The effective Kondo coupling
$\lambda = iJ_K/(4dIa^d)$.

\begin{figure}[htbp]
   \centering
  \includegraphics[width=3.4in]{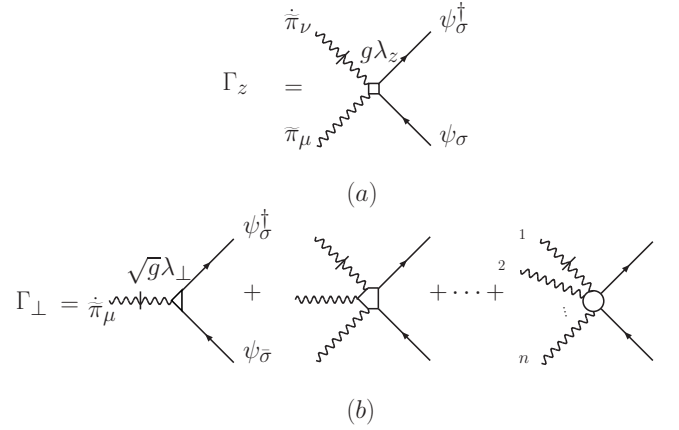} 
   \caption{The Feynman rules associate wavy lines with magnons ($\widetilde{\pi}$ fields), 
  and solid straight lines with itinerant electrons ($\psi$ fields).  A slash through a boson line indicates 
  a time derivative (i.e. $\dot{\widetilde{\pi}}$). 
(a) represents the four diagrams 
  in $\Gamma_z$. 
(b) describes the infinite number of spin-flip vertices, 
  $\Gamma_{\perp}$, involving an odd number
  of magnons.}
  \label{fig:interactionVertices}
\end{figure}

{\it Renormalization group analysis -- tree level:~}
We now carry out a renormalization group (RG) analysis of the effective 
action.
We will describe the $d=2$ case for the most part, but our
conclusions remain valid for any other $d>1$ dimensions. Our analysis involves
a combination of the bosonic RG for the 
QNL$\sigma$M~\cite{chakravarty1989,nelson1977,chaikin1995}
and the fermionic RG~\cite{shankar1994}. 
(We note in passing that a combined bosonic/fermionic RG has been used in 
the context of several other problems~\cite{pepin2004, tsai2005}.)
Without loss of generality, we take
the ultraviolet energy cutoffs for the fermions ($\Lambda_f$) and
bosons ($\Lambda_b$) to be $\Lambda \sim \Lambda_f \sim \Lambda_b$.
Unless otherwise specified, the variables $(\vec{q},\omega)$ belong to
bosonic fields, while $(\vec{K},\varepsilon)$ belong to fermionic fields,
with $\vec{K}$ measured from the Brillouin zone center
and $k \equiv K-K_F$ is relative to the Fermi
surface.
Under scaling, $\omega \rightarrow s \omega$,
$\varepsilon \rightarrow s \varepsilon$, ${\vec q} \rightarrow s {\vec q}$,
and $k \rightarrow s k$.
The fermionic kinetic term
specifies~\cite{shankar1994} that $[\psi({\vec K}, \varepsilon)]=-3/2$.

For the QNL$\sigma$M, we write $\vec{n}(\vec{x},\tau) =
\left [
\pi_+(\vec{x},\tau), \pi_-(\vec{x},\tau), \sqrt{1-\pi_+^2 - \pi_-^2}
\right ]$,
and 
define the composite vector boson field
$\vec{\varphi}$ by
\begin{eqnarray}
	\vec{\varphi}(\vec{x},\tau) &\equiv& \vec{n}(\vec{x},\tau)\times \dot{\vec{n}}(\vec{x},\tau) \nonumber \\
	&=& 
		\left(
			\begin{array}{c}
				\frac{1}{\sigma} \left(- \dot{\pi}_-
				-\dot{\pi}_+\pi_+\pi_- + \pi_+\pi_+\dot{\pi}_- \right) \\
				\frac{1}{\sigma} \left( \dot{\pi}_+ +
				\dot{\pi}_-\pi_-\pi_+ - \pi_-\pi_-\dot{\pi}_+ 
\right) \\
				\dot{\pi}_-\pi_+ - \pi_-\dot{\pi}_+\\
			\end{array}
		\right) 
\end{eqnarray}
The square-root factors can be expanded, for example 
$\frac{1}{\sigma} = \frac{1}{\sqrt{1-\pi_+^2 - \pi_-^2}} 
\approx 1+(1/2)(\pi_+^2 + \pi_-^2)+(3/8)(\pi_+^2 + \pi_-^2)^2+\cdots$.
The scaling dimensions
are
$\left[ \vec{\varphi}(\vec{x},\tau) \right] =1$ and
$\left[ \vec{\varphi}(\vec{q},\omega) \right] = -d$,
while
$[g]=1-d$. 
Note that, in order for the boson-fermion coupling term to satisfy momentum
conservation, the relative angle (which does not appear in the measure)
between $\vec{K}$ and $\vec{K}+\vec{q}$
also needs to scale~\cite{pepin2004}.
Based on all these, 
the scaling dimension of the Kondo interaction term
$\left[ d^d K \, d\varepsilon \,d^d q \,d\omega \,\psi^{\dagger}_{\alpha}(\vec{K}+\vec{q},
\varepsilon+\omega)\psi_{\beta}(\vec{K},\omega) 
\,
\vec{\varphi}
(\vec{q},\omega) 
\right]$, is $1+1+d+1+2(-3/2)-d =0 $.
We reach the important conclusion that $[\lambda]=0$: at the tree level,
the Kondo coupling is marginal in arbitrary spatial dimensions.

\begin{figure}[t]
   \centering
   \includegraphics[width=3.4in]{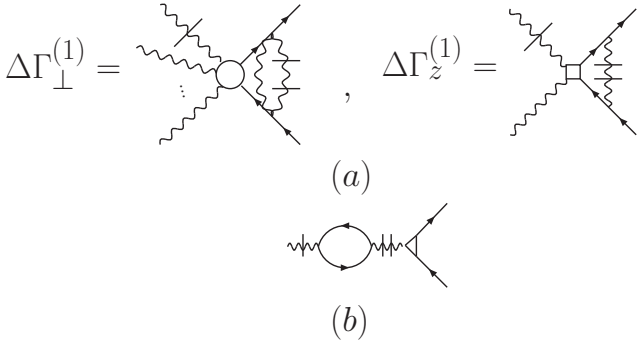}
\caption{
(a) shows the lowest order corrections to the vertices $\Gamma_z$ 
and $\Gamma_{\perp}$.
(b) is an example of a class of diagrams that do not
contribute to the beta function.
}
   \label{fig:vertexCorrections}
\end{figure}

{\it Renormalization group analysis -- one loop:~}
The Kondo interaction 
contains 
longitudinal and spin-flip terms: 
$\mathcal{S}_K 
\equiv \Gamma_z + \Gamma_{\perp}$.
It will be convenient to rescale the Goldstone field, 
$\pi = \sqrt{g}\,\widetilde{\pi}$, 
and 
the free-field 
part of the QNL$\sigma$M becomes:
$\mathcal{S}_{\text{QNL}\sigma\text{M}}
= \frac{c}{2} \int d^dxd\tau \left(\partial\widetilde{\pi}(\vec{x},\tau)\right)^2 $.
There are
an infinite number of interaction vertices involving 
an increasing number of $\widetilde{\pi}$ fields, always
coupled to exactly two fermion fields; see Fig. \ref{fig:interactionVertices}.
However, we only need to consider one representative vertex and all its loop corrections;
other vertices renormalize in the same way, as dictated by symmetry
[in a way similar to the case of the NL$\sigma$M 
itself (Ref.~\cite{chaikin1995}, p. 343)].

\begin{figure}[htbp]
   \centering
   \includegraphics[width=2.5in]{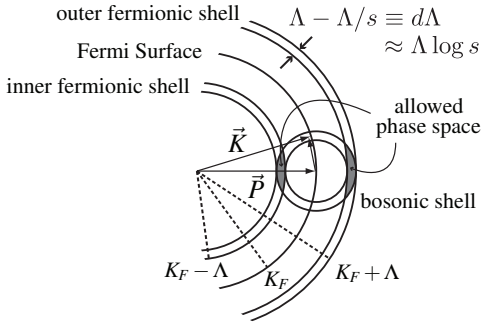} 
   \caption{
Kinematics for the momentum-shell RG. Only the shaded region is integrated over.
}
   \label{fig:phaseSpace}
\end{figure}

We describe in some detail one example of a one-loop correction, that
of $\Delta \Gamma_z^{(1)}$ shown in 
Fig.~\ref{fig:vertexCorrections}a.
Other corrections are of a similar form, and the conclusions are the same.  
Summing over the Matsubara frequency leads to
\begin{eqnarray}
	&&\Delta\Gamma_z^{(1)}(\vec{q},i\omega, \vec{P}, ip_l) = 
	g^2 \lambda_{\perp}^2 \lambda_z i\omega \times \nonumber\\
&& ~~\times (\int\limits_{\text{inner shell}} \gamma_z^{(1)}
   - \frac{1}{2}\int\limits_{\text{inner+outer shells}} \gamma_z^{(2)} \,)
\end{eqnarray}
where 
\begin{eqnarray}
\gamma_z^{(1)} &=&
	\frac{ w_{P-K}^2 (-2ip_l - i\omega +\xi_{K} + \xi_{K+q} ) }
{ \big[ (ip_l - \xi_{K})^2 - w_{P-K}^2\big] \big[(ip_l + i\omega - 
\xi_{K+q})^2 - w_{P-K}^2\big] } 
	\nonumber \\
\gamma_z^{(2)} &=&
	\frac{ 1 }{ [ ip_l- w_{P-K} - \xi_K ] }
	\frac{w_{P-K} }{ [ ip_l + i\omega - w_{P-K} -\xi_{K+q} ] } .
\end{eqnarray}
Here, 
$(\vec{P}, ip_l)$ label the energy-momentum of one of the two external fermions,
while $(\vec{q},i\omega)$ denote the energy-momentum transfer among
the two external bosons (or, equivalently, that of the two external fermions).
The magnon energy is $w_{P-K} = c|\vec{P}-\vec{K}|$.

We can now consider the kinematics of these one-loop corrections.
Three momenta,
$\vec{P}$, $\vec{K}$, and $\vec{q}$, 
are 
involved in the integrals for $\Delta\Gamma_z^{(1)}$.  
The external-fermion momentum $\vec{P}$ 
can be set to the Fermi momentum, $|\vec{P}| \approx K_F$, since 
any difference would be irrelevant in the RG sense.
Likewise, the external-boson momentum transfer 
$\vec{q}$ can be set to zero.
The fermionic loop momentum, $\vec{K}$, is restricted to the inner and outer
shells straddling
the Fermi surface:
$K_F+\Lambda/s < |\vec{K}| < K_F + \Lambda$ and
$K_F - \Lambda < |\vec{K}| < K_F - \Lambda/s$, respectively.
Finally, the bosonic momentum $\vec{P}-\vec{K}$ must be contained inside
the circle defined by its cutoff $\Lambda$.  

These restrictions on $\vec{P}$ and $\vec{K}$ lead to the construction shown in Fig. \ref{fig:phaseSpace}.
The only phase space allowed by momentum conservation is the 
shaded region in the figure.  This limits
the loop integration over $\vec{K}$ to the small angular interval from $-\Lambda/K_F$ to $+\Lambda/K_F$,
and two radial shells of width $d\Lambda \equiv \Lambda - \Lambda/s \approx \Lambda\log s$ 
(where $s \gtrapprox 1$).
A simple geometric analysis shows that 
the allowed phase space (shaded region) 
is proportional to $\Lambda^2 (\log s)^{3/2}$, therefore $\Delta\Gamma_z \propto (\log s)^{3/2}$.
The vertex correction is superlinear in $\log s$, so \textit{it does not 
contribute to the
beta function}!  
The 
Kondo coupling is still marginal at the one-loop level.

We note that if, instead of eliminating modes within the momentum shell, we integrate
over the entire phase space, the vertex correction is of the order 
$g^2\lambda_{\perp}^2\lambda_z \frac{\Lambda}{K_F}$. 
This is consistent with the vanishing contribution to the beta function
in the low-energy limit.

Finally, there are also vertex corrections due 
to the interactions purely
among the fermion fields or
purely among the QNL$\sigma$M fields.
The former do not yield loop corrections
in the forward-scattering 
channel~\cite{shankar1994}.
The latter are irrelevant since $g$ renormalizes to $0$. 

{\it Renormalization group analysis -- to infinite loops:~}
The kinematic arguments so far are similar to what happens to the 
renormalization of the forward-scattering interactions in the
pure fermion problem, where momentum conservation combined with
cutoff considerations severely limit the available
phase space \cite{shankar1994}.
The parallel
carries over to the RG beyond one loop.
We decompose the Fermi surface into $N_{\Lambda} \equiv 
\pi K_F/\Lambda$ patches, and rescale the momentum and energy variables
for each patch in terms of $\Lambda$: 
$\bar{\varepsilon}=\varepsilon/\Lambda$ and so on.
We also absorb a factor $\Lambda^2$ into the fermion field,
so that the kinetic term for the fermions becomes 
$\sum_{i}^{N_{\Lambda}} \int d^2\bar{k}_i d\bar{\varepsilon}_i
\psi^{\dagger}_i (i \bar{\varepsilon}_i-v_F 
\bar{k}_i)\psi_i$. Likewise, we absorb a factor $\Lambda^{5/2}$
into the $\widetilde{\pi}$ field, so that the kinetic part 
of the QNL$\sigma$M
is 
$\mathcal{S}_{\text{QNL}\sigma\text{M}}
\sim \int d^2\bar{q} d\bar{\omega}
(\bar{q}^2 + \bar{\omega}^2) \widetilde{\pi}^2$.
We then find that the spin-flip Kondo coupling ($\Gamma_{\perp}$)
contains a factor $\Lambda^{1/2}$, and the longitudinal
Kondo coupling ($\Gamma_{z}$) contains a factor $\Lambda$.
In other words, the Kondo couplings are of the order of 
$(1/\sqrt{N_{\Lambda}}) 
\lambda_{\perp} 
\sum_i
\int 
\varphi
\psi^{\dagger} \psi
$
and $(1/N_{\Lambda})
 \lambda_z  \sum_i
\int \varphi \psi^{\dagger} \psi
$, respectively.
These extra $1/\sqrt{N_{\Lambda}}$
and $1/N_{\Lambda}$ factors make their 
contributions negligible to infinite loops,
except for a chain of particle-hole bubbles (in the spin-flip channel),
the lowest order of 
which is shown in 
Fig.~\ref{fig:vertexCorrections}b.
The latter does not contribute to the beta
function, since the two conduction electron poles are located 
on the same side of the real axis~\cite{shankar1994}.

The Kondo coupling is therefore marginal to infinite loops.
This contrasts to what happens 
in the single-impurity
Kondo problem. 
There,
the 
Kondo coupling is 
relevant,
and flows to infinity, 
which signifies 
singlet formation in the ground state and a concomitant 
Kondo resonance in the
excitation spectrum.
In the paramagnetic phase of the Kondo lattice, the Kondo coupling 
is believed to flow to
a related strong coupling fixed point where, again, 
Kondo resonances are generated and the Fermi
surface becomes large.

In our case,
a marginal Kondo coupling implies that
there
is no Kondo singlet formation and the Fermi surface will stay small
in the sense defined earlier.

{\it Large N limit:~}To see explicitly the small Fermi surface, we turn to
a large N generalization (this is different from the previous $N_{\Lambda}$)
 of the effective action~\cite{yamamoto2006}.
The $N \rightarrow \infty$ limit
is taken with the spin symmetry of the conduction electrons enlarged 
from SU(2) to SU(N), and the symmetry of the magnons from 
O(2)
to O(N$^2$-2). 
The effective Kondo coupling is rescaled to $\lambda/\sqrt{N}$.
Leaving details for elsewhere~\cite{yamamoto2006},
we quote the equation for the 
conduction-electron self-energy,
$\Sigma(\vec{K},\tau) = 
\int d \vec{q} 
\lambda^2 G_{\varphi,0}(\vec{q},-\tau)
G(\vec{K}+\vec{q},\tau)$, where
$G_{\varphi,0}(\vec{q},-\tau) = 
\langle T_{\tau} \dot{\widetilde{\pi}}(\vec{q},-\tau)
\dot{\widetilde{\pi}}(\vec{q},0)
\rangle _{QNL\sigma M}$,
and 
$G (\vec{K},\tau)=-\langle T_{\tau} c(\vec{K},\tau) c^{\dagger}(\vec{K},0)
\rangle$ is the full
conduction-electron propagator.
We find that
$\Sigma (\vec{K},\omega)=a\omega - i b |\omega|^d {\rm sgn} \omega$,
where $a$,$b$ are constants whose dependence on $\lambda$ has the
first non-vanishing term $\sim \lambda^2$. 
(When the four-fermion interaction among the $\psi$'s is included,
there will be an $\omega^2$ term added to ${\rm Im}\Sigma$.)
It follows from $G (\vec{K},\omega) = [\omega - \xi_{\vec{K}} - 
\Sigma(\vec{K},\omega) ]^{-1}$ that the Fermi surface is the same as that 
of the conduction-electron component alone. The Fermi surface is
indeed small.


We now turn from the asymptotically exact results to their implications.
It is well accepted that two other phases specified earlier 
occur in the zero-temperature phase
diagram of the Kondo lattice: a paramagnetic phase with a large Fermi surface,
${\rm PM_L}$, and an antiferromagnetic one with a larger Fermi surface, ${\rm AF_L}$. 
The existence of ${\rm  PM_L}$ has been most explicitly seen in the
large-$N$ limit of the SU(N) generalization of the 
model~\cite{auerbach1986, millis1987}
[where $\Sigma(\vec{K},\omega) = (v^*)^2/(\omega-\epsilon_f^*)$ contains
a pole and, correspondingly,
$G(\vec{K},\omega)$ yields a large Fermi surface].
Our results here demonstrate that the antiferromagnetic part of the phase
diagram in principle accommodates a genuine phase transition from 
${\rm AF_S}$ to ${\rm AF_L}$. For commensurate antiferromagnetic ordering
(and to order $\Delta/W$ in the incommensurate case, where
$\Delta$ is the SDW gap of the ${\rm AF_L}$ phase), this 
corresponds to a Lifshitz transition
with a change of Fermi surface topology. Such a transition has been heuristically
discussed in the past~\cite{si2003, senthil2004a}; our exact result
on the stability of the ${\rm AF_S}$ phase provides
evidence for the existence of this Lifshitz transition.

In addition, the existence of the ${\rm AF_S}$ phase 
opens the possibility for a direct
quantum transition from the ${\rm AF_S}$ to the ${\rm PM_L}$ phases. 
For this transition to be 
continuous, the quasiparticle residues $z_S$ and $z_L$ must vanish when
the QCP is approached from the two respective sides.
The quantum critical point 
is then
a non-Fermi
liquid with a divergent effective mass;
local quantum criticality~\cite{si2001, si2003}
is one such example. The results reported here,
therefore,
provide a new perspective to 
view the local quantum criticality.


{\bf Acknowledgments:} 
We thank R. Shankar for an instructive discussion on the fermionic RG,
E. Abrahams, C. Bolech,
A. V. Chubukov, E. Fradkin, S. Kirchner, 
A. Ludwig, A. Rosch, S. Sachdev, T. Senthil
and F. Steglich for useful
discussions, 
NSF,
the Robert A. Welch Foundation
and the W. M. Keck Foundation
for partial support,
and 
KITP, ACP, and the Lorentz Center
for hospitality.

\end{document}